\def\GHz{{\rm GHz}}

\def\beq#1{\begin{equation}\label{#1}}
\def\eeq{\end{equation}}
\def\beqa#1{\begin{eqnarray}\label{#1}}
\def\eeqa{\end{eqnarray}}

\def\spose#1{\hbox to 0pt{#1\hss}}
\def\simlt{\mathrel{\spose{\lower 3pt\hbox{$\mathchar"218$}}
     \raise 2.0pt\hbox{$\mathchar"13C$}}}
\def\simgt{\mathrel{\spose{\lower 3pt\hbox{$\mathchar"218$}}
     \raise 2.0pt\hbox{$\mathchar"13E$}}}
\def\simpropto{\mathrel{\spose{\lower 3pt\hbox{$\mathchar"218$}}
     \raise 2.0pt\hbox{$\propto$}}}

\def\ed{\end{document}}

\def\x{\hat {\bf x}}
\def\y{\hat {\bf y}}

\def\th{\hbox{\boldmath $\theta$}} 
\def\tth{\hbox{\boldmath $\tilde \theta$}} 
\def\ep{\hbox{\boldmath $\epsilon$}} 
\def\ttheta{\hbox{$\tilde \theta$}}

\def\bi#1{\hbox{\boldmath{$#1$}}}

\documentstyle[prd,aps,epsf]{revtex}

\begin{document}
\twocolumn[\hsize\textwidth\columnwidth\hsize\csname@twocolumnfalse\endcsname




\title{The nature of the E-B decomposition of CMB polarization}
\author{Matias Zaldarriaga}
\address{Physics Department, New York University,
4 Washington Place, New York, NY 10003}
\date{May 10 2001. Submitted to Phys. Rev. D.}

\maketitle 

\vskip 1pc


\begin{abstract}

We present a derivation of the transformation between the $Q$ and $U$
Stokes parameters and the $E$ and $B$ scalar and pseudo-scalar fields.
We emphasize the geometrical properties that such transformation must
satisfy. We present the $E$ and $B$ decomposition of some simple maps
and of a model for a supernova remnant. We discuss the relative
amplitudes of the $E$ and $B$ components and argue that for generic
random maps $E$ and $B$ should have roughly the same amplitudes.

\end{abstract}


]


\section{Introduction}\label{introduction}

In the last few years there was a great surge of interest in the
polarization anisotropies of the Cosmic Microwave Background (CMB).
The detection of CMB polarization anisotropies has become a major goal
in our field, prompting many groups to build experiments and to start
thinking about future satellite missions dedicated to polarization
(see for example \cite{1998ApJ...495..580K,staggs,hedman,peterson}).

The pattern of polarization on the sky can be characterized in terms
of a scalar ($E$) and a pseudo-scalar field ($B$)
\cite{2.kks,3.spinlong}.  This decomposition is particularly useful
because density fluctuations cannot produce $B$ type polarization
\cite{3.kks,sz97}.  A $B$ type pattern is a direct signature of the
presence of a stochastic background of gravitational waves. Such
detection would provide invaluable information about Inflation (for
estimates of how constraints on parameters of the inflationary model
would improve by measuring polarization see for example
\cite{1997ApJ...488....1Z,2000ApJ...530..133T,kinney}).  This is
perhaps the most important source for the new interest in
polarization. It has also been proposed that a detection of $B$
polarization could signal other types of ``new physics'' \cite{lue}.

The mathematics of the $E-B$ decomposition has been presented in
several papers
\cite{2.kks,3.spinlong,1997PhRvD..56..596H,1998ApJ...503....1Z,tegoliv}.
In this paper we will present a different derivation of the $E-B$
transformation that will highlight the ingredients that are needed to
connect the spin two field of $Q$ and $U$ with the scalar and
pseudo-scalar fields $E$ and $B$.  The aim is to gain intuition into
the $E-B$ decomposition, which is particularly useful at this stage
when new experiments are being designed. Intuition will help address
issues such as the optimal shape of the sky patch needed to separate
$E$ from $B$, or if both $Q$ and $U$ Stokes parameters need to be
measured.

We will also present the $E$ and $B$ decomposition of some simple
polarized maps. Our aim is to understand if having a map with $B=0$,
such as the one produced by density perturbation, is something generic
or if one should always expect $E\approx B$.

The paper is organized as follow, in section \S \ref{geometry} we
present a derivation of the $E-B$ decomposition, in \S\ref{examples}
we present the decomposition for some simple intensity and
polarization maps, We comment about observational strategies and
conclude in \S \ref{discussion}.

\section{The geometrical properties of the $E-B$ decomposition}
\label{geometry}

The CMB anisotropy field is characterized by a $2\times 2$ intensity 
tensor $I_{ij}$. The intensity tensor is a
function of direction on the sky $\hat{\bf{n}}$ and  two directions
perpendicular to $\hat{\bf{n}}$ that are  used to define its components
(${\bf \hat e}_1$,${\bf \hat e}_2$).
The Stokes parameters $Q$ and $U$ are defined as
$Q=(I_{11}-I_{22})/4$ and $U=I_{12}/2$, while the temperature 
anisotropy is
given by $T=(I_{11}+I_{22})/4$ (the factor of $4$ relates fluctuations
in the intensity with those in the temperature, $I\propto T^4$). These 
three quantities fully described any state of linearly polarized light. 
While the temperature is invariant
under a rotation in the plane perpendicular to direction
$\hat{\bf{n}}$,
$Q$ and $U$ transform under rotation by an angle $\psi$ as:
\begin{eqnarray}
Q^{\prime}&=&Q\cos 2\psi  + U\sin 2\psi  \nonumber \\  
U^{\prime}&=&-Q\sin 2\psi  + U\cos 2\psi 
\label{QUtrans} 
\end{eqnarray}
where ${\bf \hat e}_1^{\prime}=\cos \psi\ {\bf \hat e}_1+\sin\psi\ 
{\bf \hat e}_2$ 
and ${\bf \hat e}_2^{\prime}=-\sin \psi\ {\bf \hat e}_1+\cos\psi\ 
{\bf \hat e}_2$. 

It is useful not to describe the polarization field in terms of $Q$
and $U$ but to do so in terms of two quantities scalar under rotation, usually
called $E$ and $B$ \cite{2.kks,3.spinlong}.  This $E-B$ decomposition is a linear
transformation of the $Q-U$ field on the sky. The transformation in
invertible. $E$ and $B$ differ in their behavior under a parity
transformation, $B$ changes sign while $E$ does not.

To make our derivation more transparent we will work in the flat sky
approximation, which is valid for small patches of sky. We do this
only for the sake of clarity as all of our results can be trivially
generalized to a full sky analysis.  In the flat sky limit the
directions (${\bf \hat e}_1$,${\bf \hat e}_2$) used to define the
Stokes parameters at every point in the plane of the sky correspond to
the coordinate axis, (${\bf \hat e}_1$,${\bf \hat e}_2$)$=(\x,\y)$.

A general linear transformation can be written as,
\beqa{lintrans0}
E(\th)&=&\int d^2\ep {\bf K}_E(\th,\ep) {\bf X}(\ep)
\nonumber \\
B(\th)&=&\int d^2\ep {\bf K}_B(\th,\ep) {\bf X}(\ep).
\eeqa
where ${\bf X}$ is the 2 component vector ${\bf X}=(Q,U)$ and ${\bf
K}_{(E,B)}$ are the transformation kernels.  

We want to derive the properties that the kernels must satisfy
to make $E$ and $B$ transform correctly. We first consider two 
types of transformations, a translation and a rotation. Under a translation
by a distance $\th_0$ 
the vectors on the plane of the sky transform as, $\th^{\prime}=\th+\th_0$. Under a
rotation of the coordinate system by and angle $\psi$ they transform as,
$\th^{\prime}= {\bf R}(\psi) \th$,
with {\bf R} the standard rotation matrix. The explicit convention for
the rotation is explained bellow equation (\ref{QUtrans}). 
In both cases $E$ and $B$ should remain
unchanged, in other words,
\beqa{lintrans}
E^{\prime}(\th^{\prime})&=&E(\th) \ \ {( \rm for \ translations}
\nonumber \\
B^{\prime}(\th^{\prime})&=&B(\th) \ \ {\rm  \& \ rotations}).
\eeqa
Equation (\ref{lintrans}) implies that,
\beqa{lintransint}
\int d^2\ep {\bf K}_E(\th^{\prime},\ep) {\bf X}^{\prime}(\ep) 
&=& \int d^2\ep {\bf K}_E(\th,\ep) {\bf X}(\ep)
\nonumber \\
\int d^2\ep {\bf K}_B(\th^{\prime},\ep) {\bf X}^{\prime}(\ep) 
&=& \int d^2\ep {\bf K}_B(\th,\ep) {\bf X}(\ep)
\eeqa
Under a translation $Q$ and $U$ are remain unchanged unchanged, 
${\bf X}^{\prime}(\th^{\prime})= {\bf X}(\th)$.
Equation (\ref{lintrans}) implies that 
${\bf K}_{(E,B)}(\th,\ep)={\bf K}_{(E,B)}(\th-\ep)$.

On the other hand, under a rotation $Q$ and $U$ are not scalars. 
They change as,
\beq{xtrans}
{\bf X}^{\prime}(\th^{\prime})= {\bf R}_X(\psi){\bf X}(\th),
\eeq
with the matrix ${\bf R}_X(\psi)$ defined in equation (\ref{QUtrans}).
Equation (\ref{xtrans}) implies that the $E-B$ decomposition
has to be non-local. $E$ and $B$ at point $\th$ cannot be constructed
by combining $Q$ and $U$ at that same point because any such linear
combination (if invertible) would not be scalar under rotations.

The other type of transformation that needs to be considered are
reflections. After a reflection $E$ remains unchanged
and $B$ changes sign,
\beqa{parity}
E^{\prime}(\th^{\prime})&=&E(\th) 
\nonumber \\
B^{\prime}(\th^{\prime})&=&-B(\th) \ \ {(\rm for \ parity) }.
\eeqa
Although equation (\ref{parity}) is valid for any reflection,
to be concrete  we consider a reflection about the $\y$ axis. The
position vectors transform as
$\th^{\prime}=(\theta_x^{\prime},\theta_y^{\prime})=
(-\theta_x,\theta_y)$ and  the Stokes parameters as
$Q^{\prime}(\th^{\prime})=Q(\th)$ and
$U^{\prime}(\th^{\prime})=-U(\th)$. The transformation laws for
reflections about other axis can be obtained by combining these
transformation laws with the transformation properties for rotations.

\begin{figure}[tb]
\centerline{\epsfxsize=9cm\epsffile{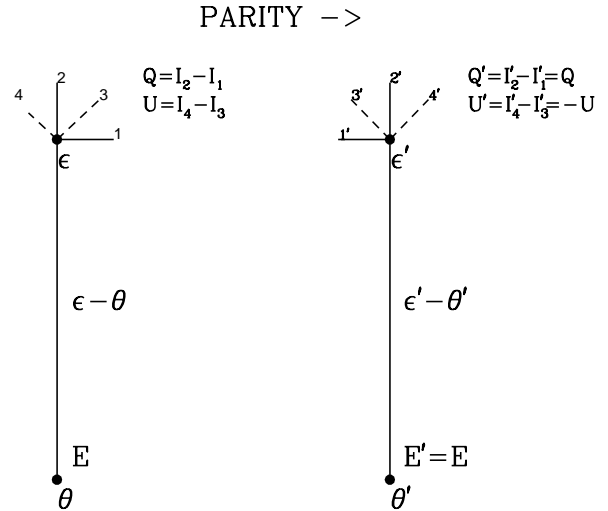}}
\caption{Contribution to $E(\th)$ from point $\ep$. The two panels are
related by a parity transformation, a reflection across the $\y$
axis. The axis labelled by numbers are the ones used to define the
Stokes parapeters. In the flat sky limit (${\bf \hat e}_1$,${\bf \hat e}_2$)$=(\x,\y)$.}
\label{fig1}
\end{figure}

\begin{figure}[tb]
\centerline{\epsfxsize=9cm\epsffile{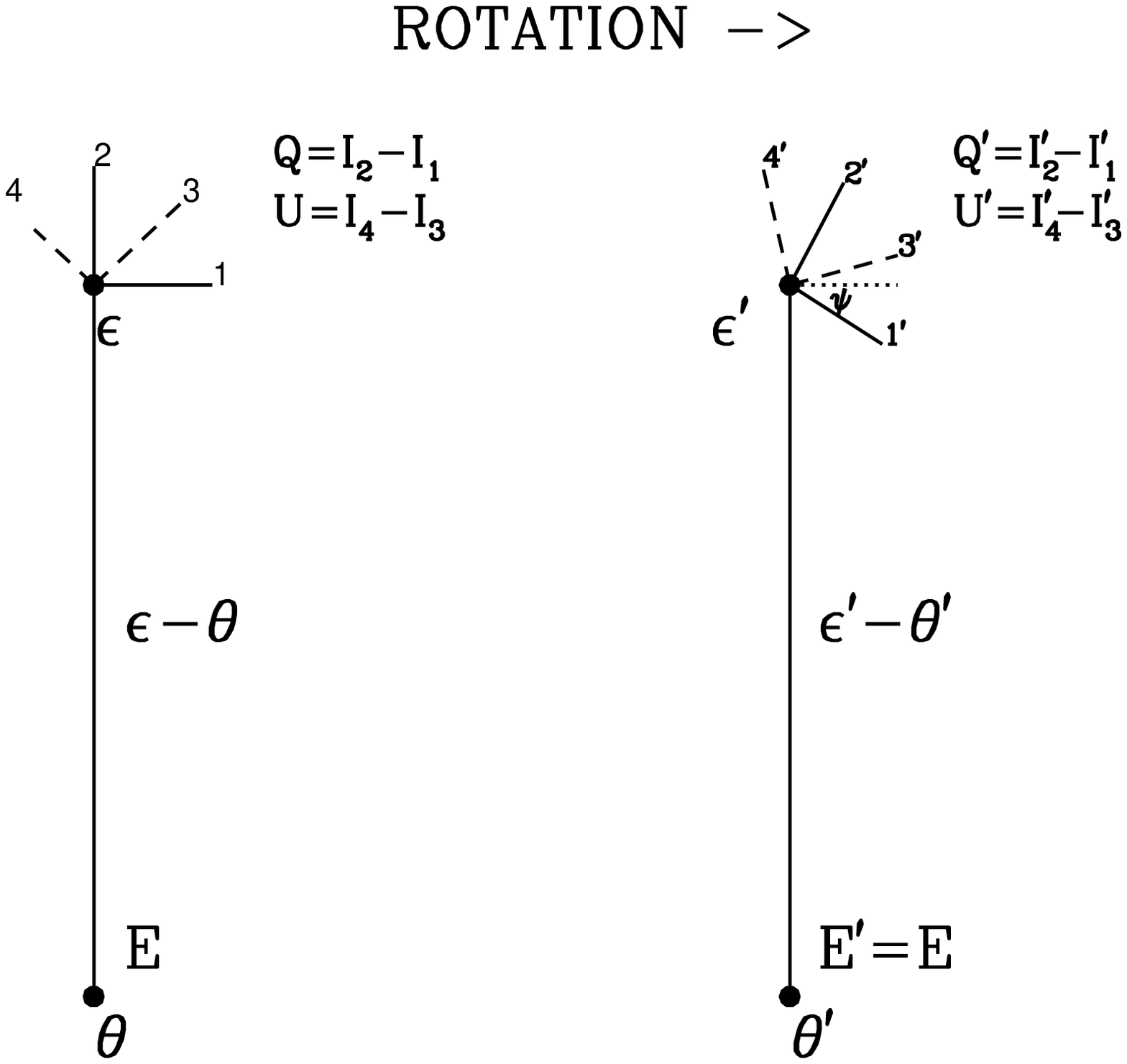}}
\caption{Contribution to $E(\th)$ from point $\ep$. The two panels are
related by a rotation by an angle $\psi$.}
\label{fig2}
\end{figure}

Rather than trying to find directly the form of the kernels needed to
satisfy all the above properties, for pedagogical reasons we will use
figures \ref{fig1} and \ref{fig2} to derive the kernels in a more
intuitive way.  We first consider the contribution to $E(\th)$ from a
point $\ep$ at a distance $\tilde \theta$ along $\y$. We will assume
that the contribution from this point in not zero. Note that with this
particular configuration the axis (${\bf \hat e}_1$,${\bf \hat e}_2$)
at point $\ep$ are aligned with the vector $\ep-\th$. This is the
reason we chose this set up. We will obtain the kernels for other
configurations using the scalar nature of $E$ under rotations.

The contribution to $E(\th)$ from 
$\ep$, which we will call $\delta E(\th)$,  is:
\beq{conte}
\delta E(\th)\propto 
\alpha_{E}^q Q(\ep)+\alpha_{E}^u U(\ep), 
\eeq 
where we have introduced ${\bf
K}_{(E,B)}(\tilde \theta \ \y)=(\alpha^q_{(E,B)},\alpha^u_{(E,B)})$.
$E$ is invariant under reflections. In figure
\ref{fig1} we consider 
a reflection across the $\ep-\th$ line, the $\y$ axis. After this
transformation $\th^{\prime}=\th$ and $\ep^{\prime}=\ep$ but
the Stokes parameters change as,
$Q^{\prime}=Q$, $U^{\prime}=-U$. This
implies that $\alpha^{E}_u=0$. 

To construct a quantity that is invariant under parity ($E$), the
contribution from a point separated by $\tilde \theta \ \y$ can only
involve $Q$. Our conclusion is a consequence of the particular
geometrical set up but we will use the transformation properties under
rotations to derive the kernels for other directions.

We use figure \ref{fig2} to understand what happens under
rotations. When we rotate the coordinate system by an angle $\psi$,
the Stokes parameter are changed as described by equation
(\ref{QUtrans}). The position angle of point $\ep$ with respect
to the $\x$ axis which we will call $\tilde\phi$ changes from $\pi/2$
to $\pi/2+\psi$. We have to allow for the kernels to depend on this
position angle otherwise $E$ would never be a scalar under rotations
given the transformation properties of $Q$ and $U$. 
Using the above argument about parity we concluded that,
\beqa{al1}
\alpha_q^{E}(\pi/2)&=& \omega(\tilde \theta) \nonumber \\
\alpha_u^{E}(\pi/2)&=& 0,
\eeqa
where we have called $\omega(\tilde \theta)$ the value of the kernel
at $\pi/2$. We have explicitly included a dependence on the separation
$\tilde \theta$ because there is no reason why all the points along
$\y$ should contribute equally. $E$ is a scalar so
\beqa{al2}
\delta E^{\prime}(\th^{\prime})&=&\delta E(\th) \nonumber \\
\alpha_E^q(\pi/2+\psi) Q^{\prime}+
\alpha_E^u(\pi/2+\psi) U^{\prime}&=&\omega Q.
\eeqa
This equation should be valid for arbitrary values $Q$ and $U$. 
Combining equation (\ref{QUtrans}) and (\ref{al2}) we obtain,
\beqa{al3}
\alpha_q^{E}(\tilde\phi)&=&-\omega(\tilde \theta) \cos(2\tilde\phi)
\nonumber \\
\alpha_u^{E}(\tilde\phi)&=& \omega(\tilde \theta) \sin(2\tilde\phi).
\eeqa

In fact equation (\ref{al3}) is simple to interpret,
only $Q_r=I_r-I_t$, the difference
between the radial and tangential intensities can be used to construct
$E$ and the weight can only depend on the distance $\tilde \theta$. 
In other words when constructing $E$ at $\th$ we should use 
the radial and tangential unit vectors to define the Stokes parameters,
(${\bf \hat e}_1$,${\bf \hat e}_2$)$=$(${\bf \hat e}_r$,${\bf \hat e}_\phi$).
In this frame only $Q_r$ contributes to $E$.

We have proven that $E(\th)$ can we expressed as,
\beqa{Efinal}
E(\th)&=&\int d^2 \tth\ \omega(\tilde \theta) Q_r(\th+\tth)
\nonumber \\
&=&\int d^2\tth \
\omega(\tilde \theta)\ [Q(\th + \tth)
\cos(2\tilde \phi) \nonumber \\
&-& U(\th + \tth)
\sin(2\tilde \phi)].
\label{Econv}
\eeqa
Any choice of weight $w$ will produce a quantity that is
scalar under rotation and invariant under parity. 

A similar argument  can be used to show that the only 
way to construct $B$, a quantity invariant under rotations but 
that changes sign under reflections is,
\beqa{Bfinal}
B(\th)&=&\int d^2\tth\ \omega(\tilde \theta) U_r
\nonumber \\
&=&\int d^2\tth \
\omega(\tilde \theta)\ [Q(\th + \tth)
\sin(2\tilde\phi) \nonumber \\
 &+& U(\th + \tth)
\cos(2\tilde\phi)]
\label{Bconv}
\eeqa
where now $U_r$ is the $U$ Stokes parameter defined with respect to
the radial and tangential directions. In general $\omega$ in equations 
(\ref{Econv}) and (\ref{Bconv}) need not be the same.

The usual definition of $E$ and $B$ corresponds 
to a particular choice of the weight $\omega$,
\beq{w}
\omega({\tilde \theta})=-1/{\tilde \theta}^2 \ \ ({\tilde \theta} > 0),
\eeq
($\omega(0)=0$ but as will become apparent later 
this fact is not important for smooth fields).
There are several reasons why this choice is made, some of which are
easier to understand when working in Fourier space. As equations 
(\ref{Efinal}) and (\ref{Bfinal}) make clear, the $E-B$ transformation
is a convolution and thus becomes a multiplication in Fourier
space. The choice of $\omega$ is such that the relation
between $E-B$ and $Q-U$ is a simple rotation in Fourier space with no
scale dependent factor. With $\omega$ given in equation (\ref{w}) the
relation is,
\beqa{FouRel}
Q(\bi{l})&=&[E(\bi{l}) \cos(2\phi_{l})
-B(\bi{l}) \sin(2\phi_{l})]
\nonumber \\
U(\bi{l})&=&[E(\bi{l}) \sin(2\phi_{l})
+B(\bi{l}) \cos(2\phi_{l})].
\eeqa
Furthermore with this choice the ensemble average of 
$P=Q^2+U^2$ is the same as the ensemble average of $E^2+B^2$. 
The sign convention on the other hand is chosen so that positive values
of $E$ generate a tangential pattern of polarization. The convention
is rooted in the weak lensing literature which has identical
mathematics and where the $E$ field corresponds to the projected
density $\kappa$ which produces tangential distortions when positive. 

For the purpose of finding a linear combination of $Q-U$ that tests
for the presence of gravitational waves or $B$ type polarization any
choice of $\omega$ is equally good. Other practical considerations
such as the geometry of the observed patch of sky will probably be
more important. In weak lensing there is a long literature that deals
with different choices of $\omega$, to create for example measures of
the enclosed mass that are more local than the $1/\theta^2$
weighting. It is beyond the scope of this section to summarize that
literature.

We want the reader to take away three basic points from the above exercise:
\begin{itemize}
\item The construction of $E$ and $B$ out of $Q$ and $U$ is by
its very nature non-local. 
\item To construct scalars under rotation at point $\th$ we need to
average the Stokes parameters around circles centered at $\th$ using
the radial and tangential directions of this circle to define the
Stokes parameters $(Q_r,U_r)$.  The weight along the circle should be
constant.
\item To construct $E$ (a scalar) we need to average $Q_r$ 
and to construct $B$ (a pseudo-scalar) we need to average $U_r$. 
\end{itemize}

\section{Examples}\label{examples}

In this section we consider simple intensity-polarization maps to
illustrate some of the properties of the $E-B$ decomposition. As a 
byproduct we will understand better if the fact that density
perturbations produce only $E$ type polarization is a unique
prediction or if most sources of polarized emission have this
characteristic.

\subsection{Simple Maps}

We start by considering a localized source of radiation of typical extent 
$L$ and centered around the origin.  We first intend to compute
the $E-B$ fields for points far away from this distribution. 
In this limit, equation (\ref{Efinal}) and (\ref{Bfinal}) become,
\beqa{EBfar}
E(\th)&=& {\cos(2 \phi) \over \theta^2} \int d^2\tth Q(\tth) -
{\sin(2 \phi) \over \theta^2} \int d^2\tth U(\th)
\nonumber \\
&& \\
B(\th)&=& {\sin(2 \phi) \over \theta^2} \int d^2\tth Q(\tth) +
{\cos(2 \phi) \over \theta^2} \int d^2\tth U(\tth). \nonumber
\eeqa

The choice of $\omega$ implies that the amplitude of $E$ and $B$
decay as $1/\theta^2$. What is more interesting is that whether $E$
or $B$ are different from zero depends on the direction of
observation. In fact there is no way to make $B$ zero everywhere and
keep $E$ different from zero. $E$ and $B$ are only zero everywhere if
the source is not polarized on average. 

In the case where the average polarization is along the $\x$ or
$\y$ axis, $E$ and $B$ are,
\beqa{EBfar2}
E(\th)&=& {\cos(2 \phi) \over \theta^2} \bar Q \ \Omega_s
\nonumber \\
B(\th)&=& {\sin(2 \phi) \over \theta^2} \bar Q \ \Omega_s
\eeqa
where $\bar Q$ gives the average polarization of the source and 
$\Omega_s$ is the solid angle it subtends. The
direction dependence of $E-B$ can be easily understood in terms of
parity transformations. The mean polarization is invariant under
reflections along directions where $B=0$ and changes sign along
directions where $B\ne 0$.

As we discussed in the previous section, the $E-B$ transformation has
to be non-local. This implies that a localized source of emission will
produce $E-B$ even outside the region where $Q$ and $U$ are not
zero. Our simple exercise has shown that if the source has a mean
polarization, the typical size of the $E$ and $B$ components are the
same outside the source.

Let us now consider points inside the source. The first potential
problem is that the weight function seems to diverge at zero
distance. We consider a
smooth $Q-U$ field that can be expanded in Taylor series. We assume we
are calculating $E$ and $B$ at point $\th$ and we expand $Q$ and $U$
around that point, 
\beqa{QUexpand} 
Q(\th+\tth)&=&Q|_{\th} + {\tilde \theta_i}
Q_{,i}|_{\th} + {1\over 2} \tilde \theta_i \tilde \theta_j Q_{,ij}|_{\th}
... \nonumber \\ 
U(\th+\tth)&=&U|_{\th}+ \tilde \theta_i U_{,i}|_{\th} + {1\over 2}
\tilde \theta_i \tilde \theta_j U_{,ij}|_{\th} ...  
\eeqa 
where we denote derivatives with respect to the different 
coordinates as $\ _{,i}$ and summation of indecesis implied. 
To compute $E$ and $B$ we replace equation (\ref{QUexpand})
into equations (\ref{Efinal}) and (\ref{Bfinal}). The first terms to
contribute are the second order ones because of the $\cos(2\tilde
\phi)$ and $\sin(2\tilde \phi)$ factors in (\ref{Efinal}) and
(\ref{Bfinal}). The $1/\ttheta^2$ in $\omega(\ttheta)$ is 
canceled by the $\ttheta^2$ coming from the Taylor expansion. In fact there
is an extra $\ttheta$ from the $d^2\tth$; there is no
divergence at the origin. We also conclude that $E$ and $B$ are most sensitive to the
``quadrupole'' pattern around $\th$; the
quadratic term in the Taylor expansion.

\begin{figure}[tb]
\centerline{\epsfxsize=9cm\epsffile{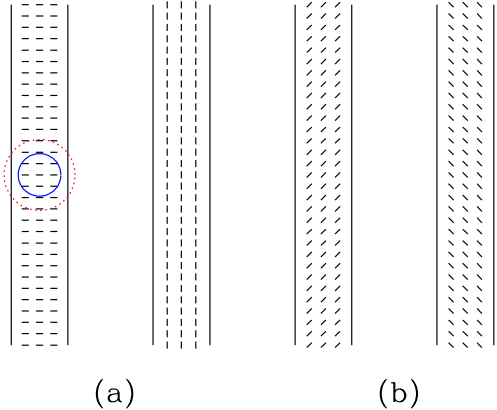}}
\caption{Examples of polarization vectors inside filaments. The two
circles indicate points that contribute with equal weight to $E$ and
$B$ at the center of those circles. The contribution from points along
the smaller circle cancel as one moves along the circle. The contribution from
the second circle is different from zero. In the case shown the
contribution is mainly $E$. The first two filaments (labelled a),
produce mainly $E$ type polarization inside the filaments while the
second two ( labelled b), produce mainly $B$ type.}
\label{fig3}
\end{figure}

As an example we can consider a filament as shown in
figure \ref{fig3}. The emitting region
has a length $L$ along the $\y$ axis and a width $l$ along the $\x$
axis, $(L>>l)$. We will assume that the Stokes parameters are constant
along the filament and are zero outside.

We now imagine doing the integrals in equations (\ref{Efinal}) and
(\ref{Bfinal}) one circle at a time. As long as the radius of the
circle is smaller than $l$ the angular integrals cancel. It is an
easy exercise to compute the values of $E$ and $B$ at the center of
the filament,
\beqa{ebcenter} 
E&=&-c\ \ Q \nonumber \\
B&=&-c\ \ U \nonumber \\
c&=&4\int_1^{L/l} dx {\sqrt{x^2-1} \over x^3},   
\eeqa
where $x=\ttheta/l$. Equation (\ref{ebcenter}) shows that $E$ and $B$
only receive contributions from rings larger than $l$ and the maximum
contribution comes from $\theta\approx \sqrt{3/2}l$. 
The contribution from far
away rings are down by the $1/\ttheta^2$. 

Although we have presented a very simplified example we see two important
features of the $E-B$ transformation. If the $Q-U$ fields are constant
over a scale $l$, rings smaller than that do not contribute to $E$ or
$B$. Around a particular point, if the polarization vectors tend to be
aligned or are perpendicular to the direction over which magnitude of the
polarization is changing the pattern has a larger $E$ than $B$. To
have $B$ the pattern has to form an angle of approximately $45^o$
with that direction. In the case of a filament this is illustrated in
figure \ref{fig3}.

\begin{figure}[tb]
\centerline{\epsfxsize=9cm\epsffile{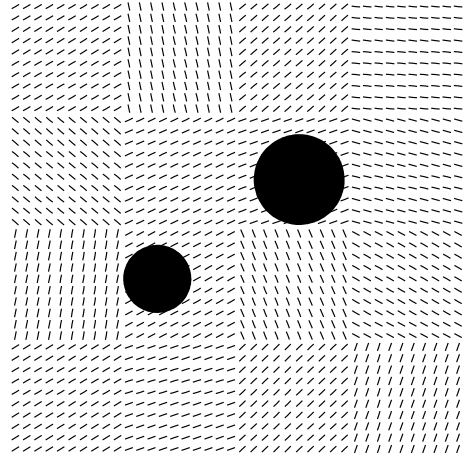}}
\caption{For illustration purposes we show a random pattern of
polarization with a coherence length we call $L$. For the two points
at the center of the filled circles, only the regions outside the circles
contribute. External patches contribute on average the same to $E$ and
$B$. Only points inside the patch (but outside the circles) will
contribute dominantly to $E$ or to $B$. Which contribution is larger 
depends on the orientation of the
polarization inside the patch and the position of the point where 
$E$ and $B$ are being calculated.}
\label{fig4}
\end{figure}

Finally we consider a case in which the polarization patter is
very random, formed by regions of finite extent of typical size
$L$ inside which the polarization is constant.  Different patches are
independent. We sketch such a pattern in figure \ref{fig4}.  We
want to know how the typical values $E$ and $B$ at a point $\th$
inside a particular region compare. From our above examples we can
conclude that the contributions to $E$ and $B$ coming from external
patches are statistically the same, only depending for any particular
external patch, on the relative orientation of the polarization in
that patch and the separation vector $\tth$. The contribution from
points inside the same patch cancels to a great extent but some
$E$ and $B$ are left. Which dominates at a particular $\th$ depends
again on the relative orientation of the polarization and the
separation vector between $\th$ and the center of the patch. Thus
for a random pattern we expect similar levels of $E$ and $B$.

Our final example argues that for polarization patterns that have a
finite coherence length one should expect to have roughly the same $E$
and $B$. This shows how remarkable it is that density perturbations do
not produce any $B$ modes. In order for $B$ to be zero, the integral of
$U_r$ has to vanish identically (not just statistically) for every
possible ring around any point, regardless of the radius of the
ring. It is clear that this important symmetry will not hold for most
random processes.

\subsection{Supernova remnant}

In this section we consider a more realistic model for polarized
emission, a model for the emission of a supernova remnant (SNR)
\cite{snm}. It has been used successfully to model the emission
from the the Galactic Spurs at radio frequencies around 1.4 $\GHz$.

The basic features of the model can be summarized as
follows. Radiation is produced by synchrotron emission from a shell. The
thickness of the shell depends on position and (to first
order in the shell thickness) is given by: 
\beq{eps} \epsilon(\psi)= \Delta r/r= \bar \epsilon \sin(\psi) 
\eeq 
where $r$ is the radius of the shell, $\psi$ is the
angle relative to the direction of the initial magnetic field and
$\bar \epsilon$ is the maximum thickness of the shell. 

The interstellar magnetic field had a strength $B_0$ before the SN
explosion.  Later it is oriented along the surface of the shell and is
amplified to a value, $B=B_0/2\bar\epsilon$, a consequence of flux
conservation. This model assumes that energy distribution of the
particles responsible for the emission is of the form, 
\beq{ener}
N(E)dE=K E^{-\gamma}dE.  
\eeq 
Moreover it assumes equipartition
between the energy density in particles and magnetic field,
\beq{equipart} 
K \int^{E_{max}}_{E_{min}} E^{-\gamma+1} dE \sim {B^2
\over 8\pi} 
\eeq

Detailed derivations of these equations as well as parameters that can fit
different structures in our galaxy can be found in
\cite{snm,snobs}. It is not our objective to analyze what
is expected from particular structures in our galaxy but rather to
use this model to make a map of polarization and compute its $E-B$
decomposition to help build intuition. 
For this purpose the only two relevant parameters are the
angle $(\beta)$ between the plane of the sky and the unperturbed interstellar
magnetic field ($\bf B_0$) and the maximum width of the shell $\bar
\epsilon$. Without loss of generality we will assume that
$\bf B_0$ lies in the $\bf x-z$ plane (the $\bf x-y$ plane is the 
plane of the sky). The projection on the sky of the unperturbed field is
$B_0 \cos(\beta)$.

We then compute the intensity and polarization observed along each
line of sight be integrating the synchrotron emissivity along the line
of sight,
\beqa{IQU}
I &=& A\ \int dx B_{\perp}^{(\gamma+1)/2} \nonumber \\
Q &=& A\ \Pi\ \int dx B_{\perp}^{(\gamma+1)/2} \cos(2\phi) \nonumber \\
U &=& A\ \Pi\ \int dx B_{\perp}^{(\gamma+1)/2} \sin(2\phi), 
\eeqa
where $\Pi$ is the degree of polarization of the synchotrom emission, $B_{\perp}$ is the component
of the local magnetic field on the plane of the sky and $\phi$ is the
angle between ${\bf B}_{\perp}$ and the $\bf x$ axis and $A$ is a
normalization constant that depends on several parameters such as the
number density of particles.

\begin{figure}[tb]
\centerline{\epsfxsize=9cm\epsffile{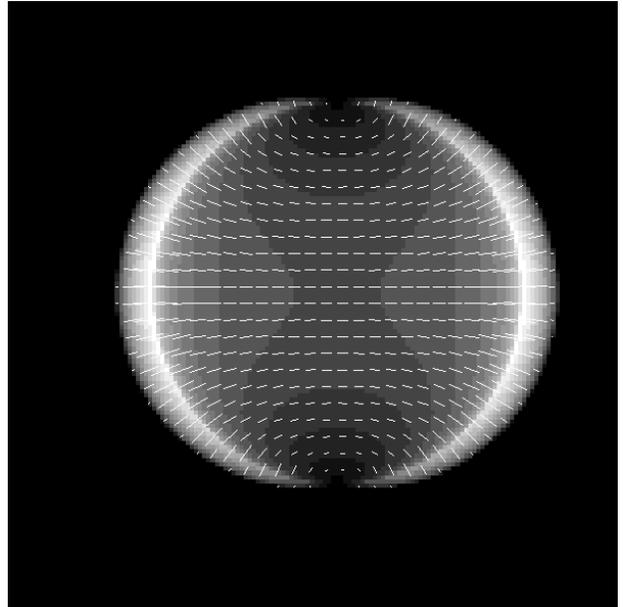}}
\caption{Temperature and polarization map for a model of SNR with
$\beta=45^o$. Rods indicate the magnitude of $P=Q^2+U^2$.}
\label{tqu}
\end{figure}

\begin{figure}[tb]

\centerline{\epsfxsize=9cm\epsffile{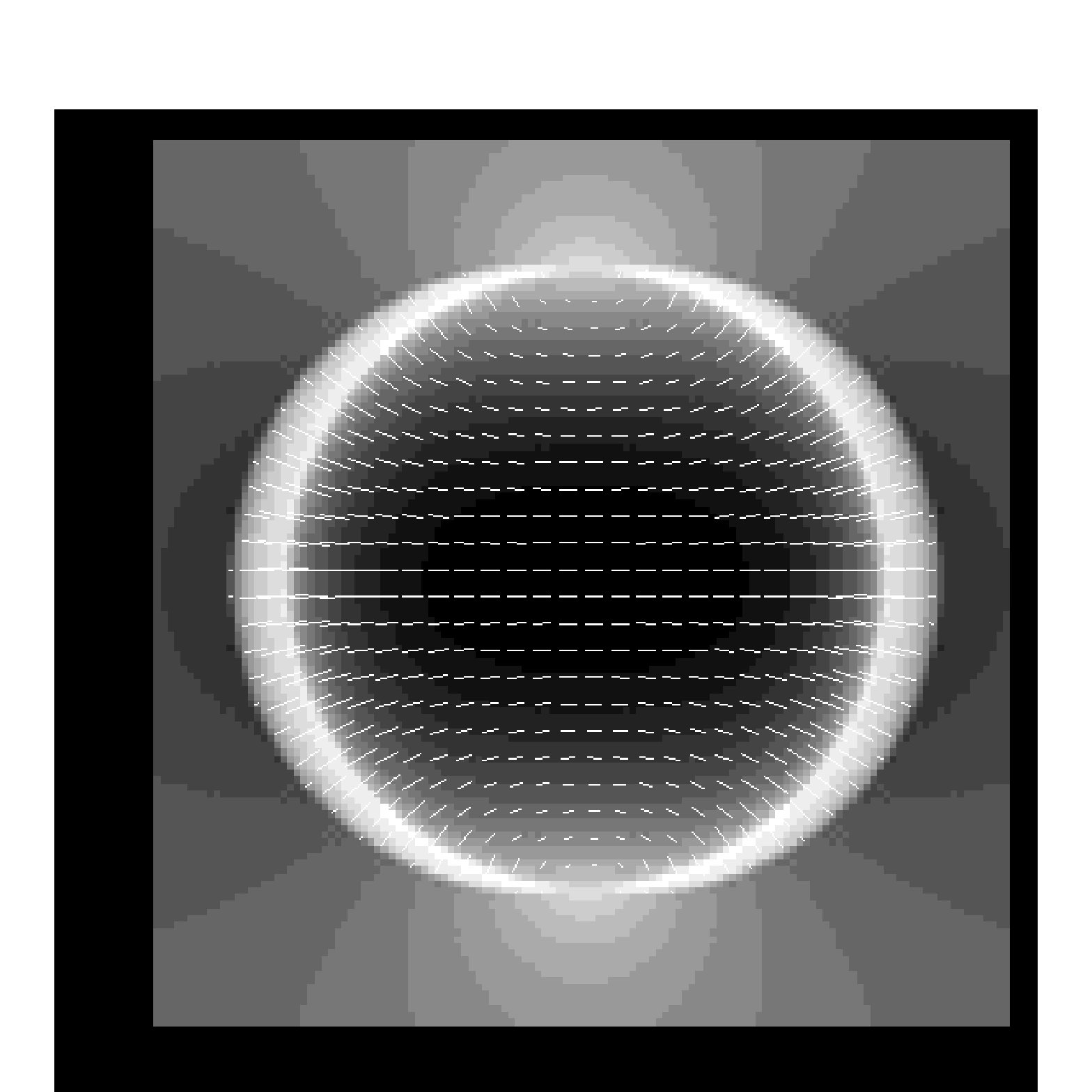}}
\caption{$E$ type polarization for a model of SNR with $\beta=45^o$}
\label{equ}
\end{figure}

\begin{figure}[tb]
\centerline{\epsfxsize=9cm\epsffile{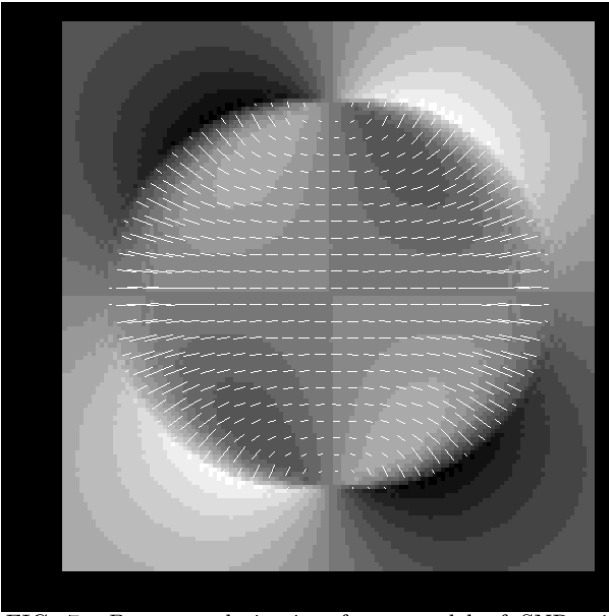}}
\caption{$B$ type polarization for a model of SNR with $\beta=45^o$}
\label{bqu}
\end{figure}

Figure \ref{tqu} shows the intensity and polarization map for the case
$\beta=45^o$. In figures \ref{equ} and \ref{bqu} we show the
corresponding $E-B$ maps. $E$ and $B$
extend outside the remnant. The intensity and polarization map has
symmetries of reflection across the $\x$ and $\y$ axis. 
The $E$ map respects those symmetries while 
the $B$ map does not. The $B$ maps changes sign as one moves across
the $\x$ and $\y$ axis.


\begin{figure}[tb]
\centerline{\epsfxsize=9cm\epsffile{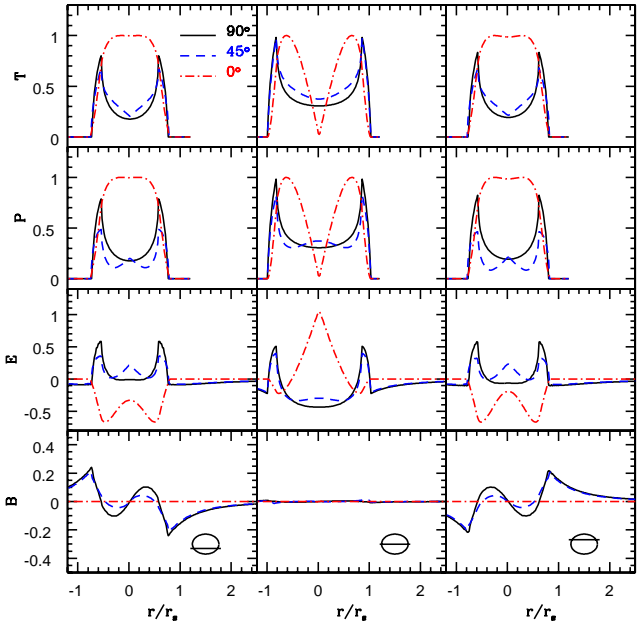}}
\caption{One dimensional cuts of $T$, $P=\sqrt{Q^2+U^2}$, $E$ and $B$
across SNR models. Each of the three columns corresponds to a
different height $y=(-0.6,0,0.6)$ in units of the radius. In each panel
we show the results for three different values of
$\beta=(0,45^o,90^o)$. The $E$ curve for $\beta=90^o$ and $y=0$ has
been divided by a factor of $10$ so all curves in the panel are
visible.}
\label{fig5}
\end{figure}

To understand the behavior of the Stokes parameters and
the $E$ and $B$ fields we will look at
one dimensional cuts along the $\x$ axis (perpendicular to the
magnetic field). We show several examples in figure \ref{fig5}. 
Each of the columns corresponds to a cut at a different height 
along the remnant.

From figure \ref{fig5} we can conclude that:
\begin{itemize}
\item Both $E$ and $B$ extend outside the remnant and decay as
$1/\theta^2$.

\item The temperature and polarization patterns are invariant under
reflections across both the $\x$ and $\y$ axis. This implies that $B$
is zero along both axis. $E$ is invariant under reflections across the
$\x$ and $\y$ axis but $B$ changes sign.

\item Both $E$ and $B$ tend to peak at the edges of the SNR, where the
$T$ and $P$ peak. We could think of the edge of the SNR as an example
of the more idealized ``filament'' that we considered in the previous
section. For $\beta=0^o$ and $45^o$ we find that $E$ is positive
at the peaks, that means that the polarization direction is
perpendicular to the direction of the ``filament''. For $\beta=90^o$
the situation is opposite and the polarization is parallel to the
filament direction.

\item $E$ is larger than $B$ by a factor of a few in a ``typical'' place 
inside the ring of maximum emission. In particular, at the edges of
the SNR, this means that the polarization is typically like patters
(a) of figure \ref{fig4} rather than patterns (b).

\item If $\beta=90^0$ $B$ is zero everywhere because the the
polarization pattern has reflection symmetry across any axis going
through the origin.

\item If $\beta=90^0$ the $E$ component has a very large peak at the
origin because the pattern is circular around that point. Note that 
there is no emission ($T=P=0$) at the origin, where $E$ has the maximum. This
illustrates the fact that one cannot define a degree of
polarization using $E$ and $B$ because their relation to $Q$ and
$U$ is not local. 
\end{itemize}

It is important to realize that many of our conclusions follow from
specific symmetry properties of the source. In reality this symmetries
will be broken by real world complications such as inhomogeneities in
the density or magnetic fields. As our examples in the previous
section show, as the symmetries are lost the amplitudes of $E$ and
$B$ become more similar. Moreover, because the $E-B$ transformation is
non-local, if only part of the SNR is in the observed field, the $E$
and $B$ decomposition will be different.
 
\section{Discussion}\label{discussion}

Detecting a $B$ component in the CMB polarization field would be a
great triumph for cosmology. As discussed above the transformation
between $Q-U$ and $E-B$ is necessarily non-local. Moreover, just from
the geometrical requirements the only way to construct scalar and
pseudo scalar quantities is to average $Q_r$ and $U_r$ over circles.  Thus
the geometry of the patch of sky to be observed should be such as to
allow for many different circles to be inscribed.

The first generation of modern experiments,
\cite{1998ApJ...495..580K,hedman} measure the Stokes parameters in a
ring on the sky. Thus only one circle can be constructed and so even
if both $Q$ and $U$ are measured at every pixel along the ring, there
is only one possible linear combination of the data that measures only
$E$ (the average of $Q_r$ along the ring) and one linear combination
that measures only $B$ (the average of $U_r$ along the ring).  All other
combinations of the data receive contributions simultaneously from
both $E$ and $B$. In practice \cite{hedman} could not use these
combinations because $1/f$ noise make them unreliable.

In general, one cannot construct circles around the points in the edge
of the observed patch. Thus if the aim is to have the most possible
linear combinations that are sensitive to either $E$ or $B$ but not to
both, the best strategy is to make the observed patch of sky as round
as possible.

Another consideration is that for each point where $E$ and $B$ are
calculated one needs to average either $Q$ or $U$ in the natural
frame, the radial and tangential directions.  This means that in order
to be able to do it for as many circles as  possibles one has to 
measure both $Q$ and $U$ in every pixel.

The emphasis of our paper was to find linear combinations of $Q$
and $U$ that contain information about $E$ or $B$ alone. However it is
perfectly possible to distinguish $E$ and $B$ type polarizations from
the correlation functions of $Q$ and $U$. Formulas that relate the
power spectrum of $E$ and $B$ with the correlation functions of $Q$
and $U$ can be found for example in \cite{2.kks}.  Distinguishing $E$
and $B$ this way does not rely on the shape of the observed
region, as the correlation functions can be calculated just using pair
of points.

One should realize however that obtaining constraints on $B$ from
correlation functions comes at the price of larger error bars. It is
clear that even though one can estimate the power spectrum this way,
the errors in an estimate of $B$ have contributions from both the
power in $E$ and $B$ type polarization. Thus, because we expect the
$B$ signal to be smaller than the $E$ one it is much better to
directly find linear combination (rather than quadratic combinations)
of the data that measure $B$. Suggestions of practical ways of
separating $E$ and $B$ from correlation data have been recently
presented in \cite{critt}. In reality a full analysis such
as the one described in \cite{tegoliv} will
incorporate all the information available in a given experiment and
should be preferred. 

We have also analyzed the $E$ and $B$ patterns expected for simple
maps. We argued that in general one expects both $E$ and $B$
type polarization to have comparable amplitudes although not
necessarily equal. Whether $E$ or $B$ dominates at a particular place
inside the source depends on the symmetries of the source. The $E$ and
$B$ transformation is not local so some $E$ and $B$ ``leaks'' outside
the source, unless on average the source is unpolarized. The amplitude
of $E$ and $B$ is the same outside the source on average but which
dominates at a particular point depends on the relative orientation
between the polarization direction and the separation vector.



\end{document}